\documentclass[aps,pra,twocolumn,amsfonts,floatfix,superscriptaddress]{revtex4}

\usepackage{graphicx}
\usepackage{dcolumn}
\usepackage{bm}
\usepackage{mathbbol}

\usepackage{amssymb}          
\usepackage{graphicx}         
\usepackage[latin1]{inputenc} 
\usepackage{amsmath}          
\usepackage{color}
\usepackage{epstopdf}

\newtheorem{prop}{Proposition}

\newtheorem{thm}{Theorem}
\newtheorem{cor}{Corollary}
\newtheorem{lemma}{Lemma}
\newtheorem{prob}{Problem}

\newtheorem{assum}{Assumption}
\newcommand{\proof}{\noindent {\bf Proof. }}

\newcommand{\ket}[1]{|#1\rangle}
\newcommand{\bra}[1]{\langle #1|}

\newcommand{\Hi}{\mathcal{H}}
\newcommand{\Si}{\mathcal{S}}
\newcommand{\Ei}{\mathcal{E}}

\newcommand{\Tr}{\mathrm{{Tr}}}
\newcommand{\supp}{\textrm{supp}}

\newcommand{\beq}{\begin{equation}}
\newcommand{\eeq}{\end{equation}}
\newcommand{\beqa}{\begin{eqnarray}}
\newcommand{\eeqa}{\end{eqnarray}}
\newcommand{\beqan}{\begin{eqnarray*}}
\newcommand{\eeqan}{\end{eqnarray*}}

\newcommand{\qed}{\hfill $\Box$ \vskip 2ex}

\newcommand{\Li}{\mathcal{L}}

\newcommand{\tr}{\textrm{trace}}

\newcommand{\ft}{\color{black}}

\begin{document}

\title{Switching Quantum Dynamics for Fast Stabilization}

\author{Pierre Scaramuzza}
\email{pierrescaramuzza@gmail.com}
\affiliation{\mbox{Dipartimento di Ingegneria dell'Informazione,
Universit\`a di Padova, via Gradenigo 6/B, 35131 Padova, Italy}}

\author{Francesco Ticozzi$^{1,}$}
\email{ticozzi@dei.unipd.it}
\affiliation{\mbox{Department of Physics and Astronomy,
Dartmouth College, 6127 Wilder Laboratory, Hanover, NH 03755, USA}}

\date{\today}

\begin{abstract}
Control strategies for  dissipative preparation of target quantum states, both pure and mixed, and subspaces are obtained by switching between a set of available semigroup generators. We show that the class of problems of interest can be recast, from a control--theoretic perspective, into a switched-stabilization problem for linear dynamics. This is attained by a suitable affine transformation of the coherence-vector representation. In particular, we propose and compare stabilizing {\em time-based} and {\em state-based} switching rules for entangled state preparation, showing that the latter not only ensure faster convergence with respect to non-switching methods, but can designed so that they retain robustness with respect to initialization, as long as the target is a pure state or a subspace.

\end{abstract}


\maketitle

\section{Introduction}

State preparation tasks are fundamental building blocks in control of experimental physical systems and quantum information protocols for quantum technologies. Quantum error correction implies preparation of a ``code'' of states \cite{ticozzi-QDS,pastawski-toric,muller,ticozzi-isometries}; measurement-based quantum computing starts from a multi-partite system prepared in an entangled state \cite{one-way}; state preparation is a benchmark for open quantum system simulations \cite{barreiro}; even a complete quantum computation algorithm, if implemented via engineered dissipation \cite{wolf-dissipation}, can be seen as a state-stabilization problem.
If the initial state is unknown or the initial state is not unitarily equivalent to the target, dissipative dynamics or measurements and feedback control must be involved in any effective preparation protocol. Existing techniques for designing such dissipative effects include feedback stabilization \cite{wiseman-book,barchielli-book,ticozzi-intro}, either measurement-based \cite{belavkin-towards,wang-wiseman,vanhandel-feedback,ticozzi-QDS,mirrahimi-feedback} or coherent \cite{lloyd-coherent,jacobs-coherent,ticozzi-cooling}, open-loop techniques based on Hamiltonian control \cite{ticozzi-NV}, and environment engineering via both open and closed loop techniques \cite{ticozzi-markovian, bolognani-arxiv, ticozzi-stochastic,ticozzi-ql,ticozzi-steady};

In this paper, we develop a new set of techniques, based on alternating continuous-time dynamics following an {\em open-loop switching} law. We shall consider piece-wise constant dynamical generators in Gorini-Kossakowskii-Sudarshan/Lindblad form. These are a particular case of generators of (two-parameter) semigroups of completely-positve, trace preserving maps. 

The idea of approximating continuous dynamics via fast switching of generators is at the basis of geometric methods for nonlinear control, and it has been successfully employed to explore controllability properties of  quantum semigroup evolutions in \cite{altafini-markovian,altafini-open,thomas-controllability,thomas-switching}.

However, our perspective is different, as the key problems we want to address concern {\em finding a switching law that asymptotically prepares a target state (or subspace), irrespective of the initial state}, as well as {\em studying which is the best performing strategy among the effective ones}. Robustness with respect to the initial state is sought after in typical experimental and information-processing tasks, as those recalled above.

Switched (and, more generally, hybrid) system stability is well studied in classical control theory \cite{liberzon-book,sun2006switched} and it is in general difficult to assess. In particular, even switching between asymptotically stable {\em linear} evolutions could lead to unstable global behavior \cite{liberzon-book}.
When quantum (one- or two-parameter) semigroup dynamics are studied as linear systems on operator spaces, two specific features become apparent: (i) no unstable behavior can emerge, since they are contractions with respect to the trace distance between operators \cite{alicki-lendi}; (ii) in order to preserve the trace of the state, they always have a zero eigenvalue, and hence admit multiple equilibria on operator space. 
%
Stability and convergence to fixed points {in Schr\"odinger's picture} is thus better studied in the so-called coherence-vector representation, which leads to an {\em affine form of the generators}. While the classical results for linear systems cannot be directly applied, we show that, conditionally to the existence of a common fixed point, a common change of representation can be constructed so that multiple affine differential equations can be transformed into linear ones. By exploiting this transformation, we are then able to derive sufficient conditions for the existence of effective, deterministic switching laws that stabilize the target. These conditions build on existing control-theoretic methods, and rely on the existence of a {\em convex combination} of the available generators that is stabilizing. 

We propose two open-loop design approaches. The first one is called {\em time-based switching}, and cycles through a set of generators approximating the stabilizing convex combination.  The other approach is called {\em state-based switching}, and requires an estimate of the initial state to compute the most convenient generator to implement. This can be done either in real time or off-line via simulation, as in model-based feedback design \cite{raccanelli-cdc}. State-based switching are shown to ensure faster convergence to the target. These methods address stabilization for both pure and mixed states in the same way, and they are easily extended to subspace stabilization.

In principle, since the second group of design methods requires the initial state to decide which generator to employ, it is in general not robust with respect to initialization. Nonetheless, we show that robustness with respect to initialization is retained {\em if the target is a pure state or subspace, and the initial estimate is full rank}. While in this case we have no guarantees on a faster convergence, numerical simulations comparing the performance of the methods indicate that the advantage persists.

The paper is structured as follows: We introduce and formalize our problem in Section \ref{sec:switching}, after reviewing basic definitions and properties of quantum Markov semigroups. Section \ref{sec:tech} illustrates our method to jointly transform the affine dynamics into linear ones, so that classical control methods of time-based switching can be applied.
The state-based methods are described in Section \ref{sec:state-based}, and relevant applications to entanglement generation in multi-qubit systems are presented in Section \ref{sec:app}. A discussion of the results and potential developments concludes our paper.

\section{Switching Quantum Markov Dynamics}\label{sec:switching}

\subsection{Preliminaries: Quantum dynamical semigroups and their stability properties}

We here recall the notation and some basic facts regarding the dynamics of open quantum systems interacting with a Markovian, yet possibly time-dependent, environment. Let $\Hi_\Si$ and $\Hi_\Ei$ denote the Hilbert spaces associated to the system of interest $\Si$ and the environment $\Ei$, with $\text{dim}(\Hi_\Si)=N<\infty$. The joint Hilbert space $\Hi_{\Si\Ei}$ is formally described by the tensor product $\Hi_\Si\otimes\Hi_\Ei$. Let ${\frak B}(\Hi)$ denote the set of linear operators on a Hilbert space $\Hi.$ A state for a quantum system is associated to a density operator, that is, a trace one, positive-semidefinite operator $\rho\in{\frak D}(\Hi)=\{\rho\in{\frak B}(\Hi)|\rho=\rho^\dag\geq 0,\;\tr(\rho)=1\}.$ We will say that the {\em support} of $\rho$ is $\Hi_\rho$ if $\textrm{range}(\rho)=\Hi_\rho.$ The evolution of a density operator $\rho$ on $\Hi_{\Si\Ei}$ is unitary, as the relative joint system plus environment is closed. Let us assume as usual that $\rho(t_0)=\rho_\Si(t_0)\otimes\rho_\Ei(t_0)$, then at each time $t\geq t_0$ the reduced dynamics on the finite-dimensional system of interest (obtained by averaging over the environment degrees of freedom by means of partial trace \cite{nielsen-chuang}), is given by a Completely-Positive, Trace-Preserving (CPTP) linear map $\Ei_{t,t_0}$.

Such kind of evolution is, in general, non-Markovian and cannot thus be described by simple dynamical equations. Nevertheless, in various cases of interest (see e.g. \cite{alicki-lendi,wiseman-book,barchielli-book}), Markovian models can be reasonably used for describing a quantum system. A dynamical system for which $\Ei_{t_2,t_0}=\Ei_{t_2,t_1}\circ\Ei_{t_1,t_0},$ for any intermediate time $t_1,$ with both the factor maps being CPTP, is said to be a two-parameter semigroup \cite{alicki-lendi, wolf-dividing}. It can be proved \cite{gorini-k-s, lindblad, alicki-lendi} that the generator of the dynamics  $\Li_t(\rho)$ can be put in symmetrized (Lindblad) form:
\begin{eqnarray}\label{eq:lindblad}
&&\hspace{-5mm}\frac{d}{dt}\rho(t)={\cal L}_t(\rho)\\&&\hspace{-5mm}=-i[H,\rho(t)]+\sum_kL_k(t)\rho(t)L_k^\dagger(t)-\frac{1}{2}\{L_k(t)^\dagger L_k(t),\rho(t)\},\nonumber
\end{eqnarray}
where $L_k(t)\in{\frak B}(\Hi)$ are called noise operators and the (effective) Hamiltonian $H$ is Hermitian. In general at most $n^2-1$ operators are needed for such identifying a generator. If the maps $\Ei_{t,t_0}$ actually depend only on the difference $t-t_0,$ the generator \eqref{eq:lindblad} is time invariant, and the one-parameter semigroup is simply called a Quantum Dynamical Semigroup (QDS) \cite{alicki-lendi}.


Let us now recall some basic notions related to invariance and stability of QDSs. A set of density operators $\mathcal{S}$ is invariant if 
$\forall \rho(t_0)\in\mathcal{S}$, $$\rho(t)=\Ei_{t,t_0}(\rho(t_0))\in\mathcal{S}, \quad \forall t\ge t_0,$$ or simply $\Ei_{t,t_0}({\cal S})\subseteq\mathcal{S}, \quad \forall t\ge t_0.$
Recall that the {\em trace distance} (or quantum total-variation distance) between two density operators $\rho$ and $\tau$ is:
\begin{equation}
\label{tracedist}
D(\rho,\sigma)=\frac{1}{2}tr[|\rho-\tau|],
\end{equation}
where the operator $|A|=\sqrt{A^\dagger A}$ denotes the positive square root of $A^\dagger A$. By defining {\em the distance of a state from a set} as
$$D(\rho,\mathcal{S}):=\inf_{\tau\in\mathcal{S}}{D(\rho,\tau)},$$
we say that a set $\mathcal{S}$ is {\em(simply) stable} if it is invariant and $\forall\varepsilon>0$ there exists $\delta>0$ such that
$D(\rho(t_0),\mathcal{S})\le\delta$ implies $\forall t\ge t_0$:
$$D(\mathcal{E}_{t,t_0}(\rho),\mathcal{S})\le\varepsilon.$$
A set $\mathcal{S}$ is {\em Globally Asymptotically Stable} (GAS) if it is simply stable and 
$\forall\rho(t_0)$ we have $$\lim_{t\rightarrow +\infty}{D(\mathcal{E}_{t,t_0}(\rho),\mathcal{S})}=0.$$

Notice that having a {\em GAS state} is equivalent to say that the QDS is {\em mixing} in the mathematical physics or Markov chains theory. We prefer the dynamical system/control theory jargon since it allows to indicate asymptotic properties of both single states and sets.

It is worth recalling two fundamental results regarding stability of QDSs: if a QDS admits a {\em unique} fixed point, then that state is GAS. In the mathematical physics literature, this is typically rephrased as ``primitivity implies mixing'' \cite{wolf-notes}. 
Furthermore, it is well-known \cite{petz-book,nielsen-chuang} that any quantum channel is a contraction in trace norm, that is 
\begin{equation}
D(\Ei_{t,t_0}(\tau),\Ei_{t,t_0}(\rho))\le D(\tau,\rho),\quad\forall t>t_0.
\end{equation}
This entails that any invariant set is automatically simply stable.
Hence, our focus will be on {\em asymptotic} stability.

\subsection{Switching dynamics and stabilization task}
\label{sdst}

In order to implement dissipative control, we assume that we are able to ``manipulate'' or engineer the environment of the system of interest in order to enact one among a set of dynamical generators, and switch between these.
More precisely, we consider evolutions driven by equations of the type \eqref{eq:lindblad} where the operators are piece-wise constant, i.e. they are allowed to change only in a countable and unbounded sequence of {\em switching times}, say $t_1,t_2,t_3\ldots$. Call ${\cal L}(t_j)$ the (constant) QDS generator of the form \eqref{eq:lindblad} active on the {\em switching interval} $[t_{j},t_{j+1}].$ Consider a generic time $t$ and let $t_k$ be the largest switching time such that $t\geq t_k.$ Since for each time-homogeneous generator the evolution map can be computed through the exponential of the generator, the global dynamics can then be written as:
\begin{eqnarray}\label{eq:switched}
\Ei_{t,t_0}&=&\Ei_{t,t_k}\circ\Ei_{t_k,t_{k-1}}\circ\ldots\circ\Ei_{t_1,t_{0}},\\
\nonumber &=&e^{{\cal L}(t_{k})(t-t_k)}e^{{\cal L}(t_{k-1})(t_k-t_{k-1})}\ldots e^{{\cal L}(t_{0})(t_1-t_{0})}.
\end{eqnarray}

\noindent We are now ready to state our state stabilization problem:

\begin{prob}\label{prob1}  Given a target set ${\cal S}\subset{\frak D}(\Hi)$, and a finite set of Lindblad dynamics $\{\mathcal{L}_j\}_{j=1}^M$ such that $e^{{\cal L}_jt}{\cal S}\subseteq{\cal S}$ for all $j$ and all $t\geq 0$, we want to find a piecewise-constant switching law $j(t),$ \[j:[0,+\infty)\rightarrow \{1,2,\ldots, M\},\] that admits a countable set of switching times $0,t_1,t_2,..$, i.e. discontinuity points, so that ${\cal S}$ is made GAS by selecting the corresponding ${\cal L}(t_k)={\cal L}_{j(t_k)}$ on $[t_k,t_{k+1})$. \end{prob}

That is, for any initial state $\tau$, \[\lim_{t\rightarrow\infty}D({\cal E}_{t,t_0}(\tau),{\cal S})=0,\]
with
${\cal E}_{t,t_0}=e^{{\cal L}_{j(t_k)}(t-t_k))}e^{{\cal L}_{k-1}(t_k-t_{k-1}))}\ldots e^{{\cal L}_{j(t_0)}(t_1-t_{0})}.$

Particular cases of interest are associated to ${\cal S}$ being a single state $\bar\rho$, or ${\cal S}$ being the set of states that have support only on a given subspace $\Hi'\subset\Hi_{\cal S}.$ 
Of course, the main problem will be determining {\em when} such a problem admits a viable solution, depending on the available generators.

Consider the case in which the target consists in a single state $\bar\rho.$ A trivial case is when there is a generator for which the target $\bar\rho$ is the {\em unique} invariant state. By the results we recalled before, the target is made GAS by simply selecting that generator for the whole evolution. An analogous argument holds for subspaces \cite{ticozzi-QDS}.
The interesting and more challenging case
is when {\em all} generators have the target $\bar\rho$ as a fixed point, but {\em each} generator also has other fixed points. Clearly, if there were more than one common fixed point to all generators, stabilization could not be achieved. A {\em sufficient condition} for asymptotic stabilizability is provided in the next Section (Assumption \ref{ass1}).

\section{Joint linear representation and Time-Based Switching}\label{sec:tech}

\subsection{Coherence-vector representation}
In order to better adapt classical switching systems techniques to the quantum framework, it is useful to employ their vectorized form. We can univoquely associate a $N^2$-dimensional vector $v_\rho$ to a $N\times N$-dimensional density operator $\rho$ by choosing a basis of $N\times N$ matrices. We choose the so-called {coherence vector representation}: we consider the opportunely scaled identity matrix $F_0=\frac{1}{N}I_N$ and the set of $N^2-1$ extended Gell-Mann matrices  $\{F_j\}$ as an orthogonal basis of the Hilbert space of Hermitian matrices \cite{alicki-lendi}.  The components of $v_\rho$ can be computed as the (Hilbert-Schmidt) inner product $v_{\rho,j}=\tr{(\rho F_j)}.$ Thus, the first component of $v_\rho$ is the same for each N-dimensional density operator $\rho$ as the trace is constant:
\begin{equation}
v_0=\frac{1}{{N}}\Tr(\rho I_N)=1.
\end{equation}
The dynamics are associated to the vectorized generator, that is a $N^2\times N^2$ real matrix whose elements are
\begin{equation}
\{\hat{L}\}_{jk}=\Tr(\Li(F_j)F_k^\dag), \qquad j,k=0,1,\dots N^2-1.
\end{equation}
This is sometimes called the ``super-operator'' form of the generator. Considering that the trace of a density operator is a constant of motion, the linearized form of \eqref{eq:lindblad} reads:\begin{equation} 
\dot{v}_\rho=\hat{L}v_\rho=
\left[\begin{array}{c|ccc}
0 & 0 & \ldots & 0 \\ \hline \\
b &  & A  & \\ \\
\end{array}\right]\left[\begin{array}{c}
1 \\ \hline \\
v_r  \\ \\
\end{array}\right]
\end{equation}
If we consider only the evolution of the {\em coherence vector} $v_r$, we obtain an affine representation of the dynamics:
\begin{equation}
\label{affine}
\dot{v}_r=Av_r+b.
\end{equation}

\subsection{The reference case: Unital generators}

Let us suppose, for now, that all the generators $\Li_j$ are unital, {i.e.} they all share the {\ft$\rho=\frac{1}{N}I$} as a fixed state. In this case, all the affine components $b_j$ of  \eqref{affine} are null and the coherence vector dynamics are therefore linear. {\ft They can be represented as:
\begin{equation}
\label{linear}
\dot{v}_r=A_jv_r.
\end{equation}
Switching linear systems in the classical case have been widely studied and a wide spectrum of results are available. 

Recall  that a square matrix $A$ is said to be {\em Hurwitz (or stable, or asymptotically stable)} if its eigenvalues have all strictly negative real part \cite{khalil-nonlinear}, and {\em marginally stable} if they have non-positive real parts, and some real part is zero with corresponding multiplicity one. It is then easy to show that an {\em autonomous, time invariant} linear system $\dot x =Ax$ admits $x=0$ as a GAS state if and only if $A$ is Hurwitz. We introduce next a typical assumption on the switching linear systems associated to matrices $\{A_j\}$ that ensures stabilizability, and that will also represent the cornerstone of most of our results. }
\begin{assum}
\label{ass1}
There exists a convex combination $A_c$ of matrices $A_j$ which is Hurwitz, that is
\begin{equation}\label{hyp1}
\exists \hspace{2mm} \alpha_1,\hdots,\alpha_m \quad s.t \quad \sum_{j=1}^m\alpha_jA_j=A_c, \quad \sum_{j=1}^m\alpha_j=1.
\end{equation}
and $A_c$ is Hurwitz.
\end{assum}
It is proved \cite{wicks1994construction, wicks1998switched, sun2006switched} that if the $A_j's$ are the generators of linear dynamics, this assumption gives a sufficient condition for the existence of stabilizing switching rules. Let $\mod(t,\varepsilon)$ be the remainder of the division of $t$ by $\varepsilon,$ namely we have $t=q\times \varepsilon+ \mod(t,\varepsilon)$ where $q$ is the integer part of $t/\varepsilon.$
\begin{prop}\label{prop:linswitch}
Suppose that Assumption \ref{ass1} holds for some $\{\alpha_j\}$. Then $v=0$ can be made GAS for the corresponding switching system \eqref{linear} for sufficiently short switching times. In particular, there exists $\varepsilon>0$ such that the switching law
\begin{equation}
\label{stabilswitch1}
\sigma(t)=\begin{cases} 1 & \textrm{if} \mod(t,\varepsilon)\in[0,\alpha_1\varepsilon) \\
2 & \textrm{if} \mod(t,\varepsilon)\in[\alpha_1\varepsilon,\alpha_1\varepsilon+\alpha_2\varepsilon) \\ \vdots & \\ m &  \textrm{if} \mod(t,\varepsilon)\in[(\sum_{j=1}^{m-1}\alpha_j)\varepsilon,\varepsilon),\end{cases}
\end{equation}
makes $\rho$ GAS.
\end{prop}

The results holds for any controlled switching linear system that satisfies Assumption 1. The proof is included for completeness in Appendix \ref{app:proof1}. 

If Assumption \ref{ass1} holds, the stabilization of the completely mixed state can be easily obtained by properly switching between unital generators. However, {\em the order} of application is irrelevant, as in the first order approximation the effective generator is the weighted sum of single ones. The key point is that, on a cycle long $\varepsilon$, each $A_j$ must act for a time interval $\Delta t_j=\alpha_j\varepsilon$, where $\alpha_j$ is the coefficient of $A_j$ in the Hurwitz convex combination.

The general problem of stabilizing other states, different from the purely mixed one, requires the use of non-unital generators. Nevertheless, we will prove that this problem can be solved by {\ft formally} recasting it into the special case of unital generators. In particular we will need to find a change of basis such to make all the generators of the reduced dynamics \eqref{affine} linear.

\subsection{Joint linear representation for fixed-point stabilization}

Given a set of marginally stable Lindblad generators that all share the same fixed state $\bar{\rho}$, we would like to find a switching law that makes such state asymptotically stable. As the target state can be different from the purely mixed one, we must suppose to deal with non-unital generators $\Li_j$, whose dynamics is univocally associated to super-operators $\hat{L}_j$ and thus to an affine reduced representation
\begin{equation*}
\dot{v}_r=A_jv_r+b_j,
\end{equation*}
where $b_j\neq0$. In this case we cannot directly use the {\ft known results} for linear switching systems. Nevertheless, the following proposition allows {\ft us} to overcome this problem. 
{\ft
\begin{prop}\label{prop:T}
If all the generators $\hat{L}_j$ share the same steady state $\bar{v}=[1\quad \bar{v}_r]^T$, then there exists an invertible matrix 
\begin{equation}
T=
\left[\begin{array}{c|ccc}
1 & 0 & \ldots & 0 \\ \hline \\
T_Q &  & T_R  & \\ \\
\end{array}\right],
\end{equation}
that removes the affine component $b_j$ in \eqref{affine}, that is, that makes all the generator matrices $\hat{L}_j^d$ block diagonal:
\begin{equation}
\label{cambiobase}
\hat{L}_j^d=T
\left[\begin{array}{c|ccc}
0 & 0 & \ldots & 0 \\ \hline \\
b_j &  & A_j & \\ \\
\end{array}\right]T^{-1}=\left[\begin{array}{c|ccc}
0 & 0 & \ldots & 0 \\ \hline 0 & & \\
\vdots &  & \tilde{A}_j  & \\ 0 & & \\
\end{array}\right]. \end{equation} 
Furthermore, there exists such $T$ that satisfies:
\begin{equation}
{\color{black}T}\bar v = \left[\begin{array}{c}
1  \\ \hline \\
0 \\ \\
\end{array}\right].
\end{equation}

\end{prop}}
\proof
Let us consider the corresponding block structure of an invertible matrix:
\begin{equation}
T=
\left[\begin{array}{c|ccc}
T_S &  & T_P &  \\ \hline \\
T_Q &  & T_R  & \\ \\
\end{array}\right],
\end{equation}
where $T_S\in\mathbb{R}$, $T_P\in\mathbb{R}^{1 \times (N^2-1)}$, $T_Q\in\mathbb{R}^{(N^2-1) \times 1}$, $T_R\in\mathbb{R}^{(N^2-1) \times (N^2-1)}$.

From \eqref{cambiobase} we have
\begin{equation*}
T\hat{L}_j=\hat{L}_j^dT.
\end{equation*}
Developing the left-hand side of last equation we get
\begin{equation}
\label{dev1}
\left[\begin{array}{c|ccc}
T_S &  & T_P &  \\ \hline \\
T_Q &  & T_R  & \\ \\
\end{array}\right]
\left[\begin{array}{c|ccc}
0 & 0 & \ldots & 0 \\ \hline \\
b_j &  & A_j  & \\ \\
\end{array}\right]=
\left[\begin{array}{c|ccc}
T_Pb_j & &T_PA_j \\ \hline \\
T_Rb_j &  & T_RA_j & \\ \\
\end{array}\right],
\end{equation}
while for the right-hand side we obtain:
\begin{equation}
\label{dev2}
\left[\begin{array}{c|ccc}
0 & 0 & \ldots & 0 \\ \hline 0 & & \\
\vdots &  & \tilde{A}_j  & \\ 0 & & \\
\end{array}\right]
\left[\begin{array}{c|ccc}
T_S &  & T_P &  \\ \hline \\
T_Q &  & T_R  & \\ \\
\end{array}\right]=
\left[\begin{array}{c|ccc}
0 & 0 & \ldots & 0 \\ \hline \\
\tilde{A}_jT_Q &  & \tilde{A}_jT_R  & \\ \\
\end{array}\right].
\end{equation}
In order for \eqref{dev1} and \eqref{dev2} to be equal irrespective of $b_j, A_j$, it must be:
\begin{eqnarray}
&&T_P=0.\label{condT0}
\end{eqnarray}
Choosing for simplicity $T_S=1$ and using \eqref{condT0},
we get:
\begin{equation}
T=
\left[\begin{array}{c|ccc}
1 & 0 & \ldots & 0 \\ \hline \\
T_Q &  & T_R  & \\ \\
\end{array}\right],
\end{equation}
which is a change of basis matrix if $T_R$ is invertible. By exploiting the lower-block-triangular form of $T^{-1},$ and thus of its inverse, it is straightforward to show that
\[\hat{L}_j^d=T
\hat{L}_j T^{-1}=\left[\begin{array}{c|ccc}
0 &  & 0 &  \\ 
\hline  & & \\
T_R (b_j - A_jT_R^{-1}T_Q) &  & T_R{A}_jT_R^{-1}  & \\  & & \\
\end{array}\right].\]
We need to show that a $T_Q$ and an invertible $T_R$ exist, such that  $T_R (b_j - A_jT_R^{-1}T_Q)=0$ {\em independently of $j$. }
In order to do this, recall that $\bar{v}_r$ is a common invariant state, so that $A_j\bar{v}_r+b_j=0$ for all $j$. If we choose an invertible $T_R$ and define $T_Q=-T_R\bar{v}_r,$ we have
\[b_j - A_jT_R^{-1}T_Q=b_j-A_jT_R^{-1}(-T_R\bar{v}_r)=b_j+A_j\bar{v}_r=0.\]
\qed
This change of basis allows us to transform all the affine dynamics in linear ones. {\ft It is worth remarking that, while $T$ acts linearly on the vectors $v_\rho,$ it acts as an affine transformation on their {\em coherence vectors} $v_r$. In fact, {\color{black} defining $\tilde{v}_r$ as the transformed ``coherence'' part of $v$,} we get $v_r=\frac{1}{\sqrt{N}}T_Q+T_R\tilde v_r,$ and hence:
\begin{equation}\label{eq:affinetrans}
\tilde v_r={\color{black}  T_R^{-1}v_r-\frac{1}{\sqrt{N}}T_R^{-1}T_Q }= T_R^{-1}v_r + \frac{1}{\sqrt{N}}\bar{v}_r. 
\end{equation}
}

All the transformed generators $\tilde{A}_j$ behave exactly as unital generators. It is therefore possible to apply the results we derived in the last section and the existence of a Hurwitz convex combination of all the $\{A_j\}$ is again a sufficient condition for the existence of a stabilizing switching law. By applying the time-based one described in \eqref{stabilswitch1} of Proposition \ref{prop:linswitch}, we can make the shared, {\ft {\em general} fixed state $\bar{\rho}$ asymptotically stable. The final results is summarized in
the following:
\begin{thm}\label{thm:generalstate}
Suppose to dispose of a set of generators ${\cal L}_j$ that share a common steady state $\rho$. Then there always exists an affine transformation that makes all their {\em coherence vector} representations simultaneously linear. That is, by \eqref{eq:affinetrans}, we get:
\[\dot {\tilde v}_r=\tilde A_j \tilde v_r.\] If Assumption \ref{ass1} holds for the transformed generators $\tilde A_j$, then a switching law as in \eqref{stabilswitch1} makes $\bar{v}$ asymptotically stable, for sufficiently short switching intervals.
\end{thm}
}

\subsection{Joint linear representation for subspace stabilization}

{\ft In the previous section we found a class of transformations that allow to recast the switching stabilization problem for {\em general} dynamical generators in the easier unital case, where we can use the linear system approach. Another case of interest in which it is possible to recast the problem as a {\em reduced} linear system stabilization is the asymptotic preparation of subspaces. In the notation we used in Problem \ref{prob1}, we here consider the target set ${\cal S}\subseteq {\mathfrak D}(\Hi)$ to be a set of density operators that have support on a certain subspace $\Hi_S$ of $\Hi.$ This class of problems is of key interest for initialization of error correcting and noiseless quantum codes \cite{knill-QEC,lidar-DFS, zanardi-DFS,viola-generalnoise,ticozzi-QDS}, cooling to degenerate ground subspaces \cite{ticozzi-ql,ticozzi-cooling}, and multipartite state preparation using quasi-local resources \cite{ticozzi-ql,ticozzi-steady}. Choosing an opportune basis that respects an orthogonal partition $\Hi=\Hi_S\oplus \Hi_R$, any operator can be represented as block matrix:
\beq
X=\left[
\begin{array}{c|c}
X_S & X_P \\ \hline
X_Q & X_R 
\end{array}\right]\in\mathcal{D}(\Hi).
\eeq
In order to preserve invariance of ${\cal S},$ as assumed in our Problem, any available generator ${\cal L}_j$ must satisfy \cite{ticozzi-QDS}:
\beq\label{invprop}
{\cal L}_j\biggl(\left[
\begin{array}{c|c}
\rho_S & 0 \\ \hline
0 & 0 
\end{array}\right]\biggr)=\left[
\begin{array}{c|c}
{\cal L}_{S,j}(\rho_S) & 0 \\ \hline
0 & 0 
\end{array}\right],
\eeq
for some reduced generator ${\cal L}_{S,j}$ and any density (sub)matrix $\rho_S$ on $\Hi_S.$

In order to use linear switching techniques for making $\Hi_S$ attractive, it is convenient to employ a vectorized form that explicitely separates the components related to the elements of an operator basis of ${\frak B}(\Hi)$ that have support on $\Hi_S$, and the rest. By writing ${\frak B}(\Hi)={\frak B}(\Hi_S)\oplus{\frak B}(\Hi_S)^\perp$ and considering a basis that respects such orthogonal subdivision, we obtain a block-vector representation for $\rho$:
\begin{equation}
\label{invariante}
v=\left[\begin{array}{c}
1 \\ \hline v_S \\ \hline v_S^\perp
\end{array}\right],
\end{equation}
where the upper two blocks are associated to the invariant states. The super-operator form of the generators must respect the invariance property \eqref{invprop}, namely the $\hat L_j$ , must have the following block-form:
\begin{equation}\label{eq:tristra}
\hat{L}_j=\left[\begin{array}{c|c|c}
0 & 0  & 0 \\ \hline 
b_j^S & L_j^S & L_j^X \\ \hline 
0 & 0 & L_j^\perp\\ 
\end{array}\right]\quad j=1,\dots,m.
\end{equation}
This {\em simultaneous} upper-triangular block representation of the generators will play the same role of the transformation $T$ in Proposition \ref{prop:T}, allowing us to use the same sufficient condition that we used in the unital state-preparation case.}

{\ft Our aim is to find a switching law for which $v_S^\perp$ tends to zero for ${t \rightarrow +\infty}$, for any initial state. While the dynamics of $v_S^\perp$ influences that of $v_S,$ the converse is not true: thanks to the triangular structure of \eqref{eq:tristra}, the evolution of $v_S^\perp$ is determined only by the $L_j^\perp$ of the currently selected evolution. Thus, if we assume that there exists a convex combination of $\{L_j^\perp\}$ that satisfies Assumption 1, namely
\beq
\label{assum}
\exists \hspace{0.1cm} \alpha_1\ldots\alpha_m \quad s.t \quad \sum_{j=1}^m \alpha_jL_j^\perp=L_c^\perp, \quad \sum_{j=1}^m \alpha_j=1,
\eeq
and $L_c^\perp$ is Hurwitz, then a time-based switching law can be derived in the same way as for single states stabilization. 
\begin{prop}
Let's suppose to dispose of a set of $m$ generators with a common invariant subspace and that equation \eqref{assum} holds. Then there exists a small enough $\varepsilon>0$ such that the switching law \eqref{stabilswitch1}  renders the common invariant set ${\cal S}=\{\rho\in{\mathfrak D}(\Hi)|\textrm{supp}(\rho)\subseteq\Hi_S\}$ GAS. 
\end{prop} 
}

\subsection{Switching stabilization with Hermitian generators}

In all the previous cases we need to find a Hurwitz convex combination of given matrices in order to derive a stabilizing switching law. Unfortunately, this task has been proved to be a NP-hard problem \cite{blondel1997np} in general, for which practical, scalable algorithms are unlikely to be found. However, the subclass of dynamics of interest here is easier to be studied, since they are all stable.

{\ft There is, however, a class of problems} for which it is not a necessary to find a Hurwitz convex combination of the generators. If all the generators $\hat{L}_j$ are unital, {\em have hermitian symmetry} and share only one steady state, then it is sufficient to employ each generator infinite times, for example by periodically repeating a sequence that involves all the generators. If such matrices are marginally stable, as we suppose, there exists a change of basis that shows that the euclidean norm of the coherence vector can never increase, and switching between them leads to the shared steady state. This fact is proven in the following proposition.
{\ft \begin{prop}
Consider a set of unital {\em symmetric} generators with linear representation $\hat{L}_j$, which share only one steady state $\bar{v}$. Then there exists a switching law that makes $\bar{v}$ GAS.
\end{prop}
}
\proof
According to the hypotheses, we are dealing with generators which have all the following form:
\begin{equation*}
\hat{L}_j=\left[\begin{array}{c|ccc}
0 & 0 & \ldots & 0 \\ \hline 
0 & & & \\
\vdots &  & A_j  & \\ 
0 & & & \\
\end{array}\right],
\quad A_j=(A_j)^\dag, \quad j=1,\ldots,m.
\end{equation*}
{\ft As the shared steady state is unique, it must be the completely mixed state, whose coherent part is
$$\bar{v}_r=\begin{bmatrix} 0 & \hdots & 0 \end{bmatrix}^T.$$
Being symmetric, each} matrix $A_j$ is orthogonally diagonalizable with a unitary matrix $U_j$:
\begin{equation*}
A_j^d=U_j^TA_jU_j,
\end{equation*}
where $U_j$ is an orthogonal matrix. Moreover, being marginally stable, $A_j^d$ has real negative or null eigenvalues on its diagonal. Consider the positive-definite Lyapunov function
$$V(v_r)=\frac{1}{2}v_r^Tv_r.$$
The latter is invariant for unitary change of basis, so it can also be expressed as
$$V(v_r)=\frac{1}{2}({v}_r^TU_{j}^T)(U_{j}{v}_r)=\frac{1}{2}\tilde{v}_r^T\tilde{v}_r=V(\tilde v_r),$$
{\ft where we now denote by $\tilde{v}_r(t)=U_{j}v_r(t),$ with $j$ identifying the active generator driving the evolution.} The derivative of the Lyapunov function along the systems's trajectory is
\begin{equation*}
\dot{V}(\tilde v_r)=\tilde{v}_r^T\frac{\partial\tilde{v}_r}{\partial t}=\tilde{v}_r^TA_j\tilde{v}_r\le 0,
\end{equation*}
as $A_j$ is negative semidefinite. If $\tilde{v}_r$ belongs to the eigenspace relative to an eigenvalue equal to zero, then {\ft it is left unchanged. However, if e.g. all the generators are periodically employed, as in \eqref{stabilswitch} with {\em any} convex combination with non-zero coefficients, for any $v_r\neq 0$ there exists at least one generator $A_{\hat j}$ for which
\begin{equation*}
\dot{V}(\tilde v_r)=\tilde{v}_r^T\frac{\partial\tilde{v}_r}{\partial t}=(\tilde{v}_r)^TA_j\tilde{v}_r< 0,
\end{equation*}  
for each $\tilde{v}_r$. Indeed, if this was not the case, there would be a nonzero state for which}
\begin{equation*}
\frac{\partial\tilde{v}_r}{\partial t}=A_j\tilde{v}_r=0, \qquad j=1,2,\hdots,m.
\end{equation*}
That is impossible as the origin is supposed to be the only shared steady state. Thus the norm of the reduced vector $\tilde{v}_r$ keeps decreasing and the density operator $\bar{\rho}=\frac{1}{N}I_N$ is asymptotically stable for the switching dynamics made by the Lindblad generators $\mathcal{L}_j$ that correspond to $\hat{L}_j$.
\qed    

It is worth noting that, as for Proposition \ref{prop:linswitch}, the result holds for general switching of {\em symmetric} linear systems. It is the symmetry of the generators that ensures real spectrum and thus monotone convergence to the origin. The assumption of symmetric generators entails however ``more'' robustness than in the previous case, where Assumption 1 was invoked: in fact, there is no need to consider {\em short} switching time intervals. This also implies that, for example, one could randomize the activation of the generators, instead of cycling periodically. As long as all generators appear for a total unbounded subset of times, convergence is  guaranteed. 

Of course, the same result applies when all the generators $\tilde{A}_j$ are symmetric {\em after having been transformed in a jointly-linear form.} In this case any switching law that employs all the generators enough times leads to the stabilization of the only shared steady state, which can, in general, be different from the purely mixed one. Hence, we can state this corollary of the previous proposition. 

{\ft \begin{cor}
Consider a set of generators with linear representation $\hat{L}_j$, which share only one steady state $\bar{v}$. If there exists a jointly-linear, coherence-vector representation in which $\tilde A_j=\tilde A_j^\dag$ for all $j,$ then there exists a switching law that makes $\bar{v}$ GAS.
\end{cor}}

\section{State-based switching}\label{sec:state-based}

\subsection{Preliminaries}

{\ft So far we have illustrated switching techniques for stabilizing quantum states that do not require any knowledge of the state of system, either the initial or the current ones. Hence, they are intrinsically robust with respect to initialization. Nevertheless, such switching laws do not allow for a straightforward optimization of the convergence rate to the target. This is quite clear if we consider any convex combination of two matrices describing linear systems:
$$A_c=\alpha_1A_1+\alpha_2A_2,\quad\alpha_1+\alpha_2=1.$$
If the switched system is stabilizable, there exists, in general, a range of values for $\alpha_1$ (and then for $\alpha_2$) that makes $A_c$ Hurwitz. One could in principle try to compute the {\em spectral abscissa}, i.e. the real part of the ``slowest'' eigenvalue and use that as a worst-case estimate of the asymptotic speed of convergence. However, the dependence of the spectrum on the $\alpha_i$ coefficients is in general hard to establish, and it is becoming increasingly clear that the spectral properties of evolution, also in the quantum case, may not be the only interesting performance index \cite{wolf-cutoff,wolf-bounds}.}
 
{\ft We hereby present a way to circumvent these difficulties that let us design a switching strategy with steepest descent, or at least a guaranteed minimal decrease of a quadratic function (for quantum systems, a variation of the Hilbert-Schmidt distance). In order to do that, we need to have {\em an estimation of the initial state of the system}. We can then exploit this additional information in order to find control laws that optimize convergence rates. 

The method follows closely the ideas presented in \cite{sun2006switched} for general linear systems, that we revisit in the following. As we did before, we introduce the strategy for a generic set of switching and marginally stable linear dynamics $\dot x = A_j x.$ Again, we need to suppose that Assumption \ref{ass1} holds, namely that there exists a Hurwtiz convex combination $A_c=\sum_j \alpha_j A_j$ of the linear generators of the vectorized dynamics. 

First, we associate to the asymptotically stable system generated by $A_c$ a positive-definite Lyapunov function, 
\beq
\label{defp}
V(x)=x^TPx,
\eeq 
where $P$ is the only positive solution of the Lyapunov equation, $A_c^TP+PA_c=-I$.  According to standard linear Lyapunov theory (see e.g. \cite{khalil-nonlinear}), this ensures that $V$ decreases along the trajectories generated by $\dot x=A_cx$ with rate of $-\|x\|^2.$ The function itself can be used as a pseudo-distance with respect to the target, and its decrease is a good estimate for the convergence rate.

Next, take this $P$ and, for each marginally stable $A_j$, compute the symmetric, but not strictly negative, matrices $Q_j=A_j^TP+PA_j.$ Notice that this allows to easily determine the derivative of $V$ along the trajectory driven by $A_j$, since:
\[\frac{d}{dt}V(x)=x^T(A_j^TP+PA_j)x=x^TQ_jx.\]

Then the following Lemma holds:
\begin{lemma}
\label{lemmaprec}
For each state $x\neq0$ there exists a subsystem $j$ for which
$$x(t)^TQ_jx(t)\le -x(t)^Tx(t).$$ 
\end{lemma}
\proof
As $A_c$ is a Hurwitz matrix, according to the definition \eqref{defp} of $P$,  
\begin{eqnarray*}
&&x(t)^T(A_c^TP+PA_c)x(t)\\
&&=x(t)^T\bigl((\sum_j\alpha_jA_j^T)P+P(\sum_j\alpha_jA_j)\bigr)x(t)\\
&&=\alpha_1x(t)^TQ_1x(t)+\hdots+\alpha_mx(t)^TQ_mx(t)\\
&&=-x(t)^Tx(t)<0.
\end{eqnarray*}
As the sum is negative, then there must exist at least one addend which is strictly negative too. Consider the $j$ corresponding to the smallest one. Since a convex combination of the rates $x(t)^TQ_kx(t)$ is equal to $-x(t)^Tx(t),$ then the minimum rate must be equal or smaller than the convex combination, and lemma is proved.\qed

This Lemma ensures that a generator $A_j$ that makes the Lyapunov function \eqref{defp} strictly decrease always exists, and it provides also indications on which system allows a fastest decrease of the Lyapunov function. 

\subsection{Steepest-descent switching\\ with fixed switching interval}\label{sec:steepestdesc}

As suggested by the Lemma of the previous section, a switching law that attains the fastest possible decrease rate of the Lyapunov function that is associated to the stabilizing convex combination $A_c$ can be simply obtained by selecting, {\em at each time,} the generator that ensures the steepest descent. Namely, the current value of the switching signal $j(t)$ is chosen as follows:
\begin{equation*}
j(t)=\arg\min_{k\in\mathcal{P}}\{x_t^TQ_kx_t\}.
\end{equation*}
If two or more generators correspond to the same decrease rate, we can pick either one.

In order to (locally) maximize the convergence rate, the switching signal should switch as soon as any generator allows a faster decrease of the Lyapunov function -- that is, one should compute the steepest-descent evolution at all times. 

This approach has two issues: the first one is that it may not be possible to implement such switching law -- it requires to either to be able to determine the full (continuous-time) switching law off-line, or be able to have some mapping of the state space in areas where a generator is optimal, and perform the switching when the current state, obtained by integration of the implemented dynamics, reaches a border between the areas. The second issue is that we do not know how to guarantee that a {\em finite} number of switches occurs in any finite time interval. Similarly, we do not know how to guarantee convergence if the switching time is bounded for below.
In cases where the computational task of determining the optimal switching law is not viable, or we want to prevent an infinite number of switches from occurring in a finite interval, suboptimal solutions can be devised. 

The simplest way to bypass the problems described above is to {\em impose a minimal switching time} $\Delta T>0$, and compute the steepest-descent generator at the beginning of each interval. On the one hand, this of course prevents infinite switching in any finite-length interval. On the other hand, by Lemma \ref{lemmaprec} along with the continuity of the evolution and its derivatives (apart from the switching instants), we know that there exists a small enough $\Delta T>0$ such that the Lyapunov derivative remains strictly negative for all times. This implies convergence of the steepest-descent method for some minimal switching interval. However, in some cases the latter may be to small or hard to compute. Another suboptimal strategy is described in the following section.

\subsection{Sub-optimal switching\\ with minimal guaranteed descent}
\label{sec:suboptimal}

A suboptimal solution with a guaranteed rate of convergence can be devised as follows \cite{sun2006switched}. Let us define a set of arbitrary real numbers $r_j\in(0,1]$, each of which is associated to a generator $A_j$. These are tuning parameters for the algorithm and can be taken smaller than one in order to avoid chattering and overly fast switching laws, at the price of slower convergence. They correspond to the minimal accepted descent rates for $V$.
If all $r_j$ are taken to be one, the convergence is at least as fast as the one associated to the non-switching dynamics $A_c.$ 

The algorithm is initialized as the steepest descent one: at the first instant $t_0$ a $j$ is selected such that:
\begin{equation*}
\j(t_0)=\arg\min_{j\in\mathcal{P}}\{x_{t_0}^TQ_1x_{t_0},\hdots,x_{t_0}^TQ_mx_{t_0}\}.
\end{equation*} 
The next switching instant is then chosen to be
\begin{equation*}
t_1=\text{inf}\{t>t_0: x^T(t)Q_{j(t_0)}x(t)>-r_{j(t_0)}x^T(t)x(t)\},
\end{equation*}
and the following active subsystem is that associated to
\begin{equation*}
j(t_1)=\arg\min_{j\in\mathcal{P}}\{x(t_1)^TQ_1x(t_1),\hdots,x(t_1)^TQ_mx(t_1)\}.
\end{equation*}
By Lemma \ref{lemmaprec}, we are guaranteed that such a generator ensures that the Lyapunov function strictly decreases, and does so faster than it would using the stabilizing convex combination. The sequences of switching times and active systems can be thus recursively defined:
\begin{equation}
\label{item1}
t_{k+1}=\text{inf}\{t>t_k: x^T(t)Q_{j(t_k)}x(t)>-r_{j(t_k)}x^T(t)x(t)\},
\end{equation}
and
\begin{eqnarray}
\label{item2}
j(t_{k+1})&=&\arg\min_{p\in\mathcal{P}}\{x(t_{k+1})^TQ_1x(t_{k+1}),\hdots\\ \nonumber
&&\hdots,x(t_{k+1})^TQ_mx(t_{k+1})\}.
\end{eqnarray}
Notice that in order to compute the switching law, we need the current state (either by off-line calculations, or real-time simulation -- in the spirit of model-based feedback \cite{raccanelli-cdc}). In this case, we only solve a minimum problem when the descent becomes slower than the active threshold $r_j.$ 

Given such a switching law, we need to check whether the switching system is well-posed, that is, if the set of jump times is finite for any finite interval.}
\begin{thm}\label{thm:suboptimal}
Under the above switching law, the switching system is well-posed and asymptotically stable.
\end{thm}
The proof, that follows known results for classical systems \cite{sun2006switched}, is provided in Appendix \ref{app:proof2} for completeness.

\subsection{Robustness with respect to initialization}\label{sec:robustness}

In principle, one needs to know the initial state of the quantum system of interest in order to apply state-based switching techniques. An accurate estimation of such state appears to be, in general, necessary for the successful stabilization of the target state. 

\noindent{\em Example:} Let's suppose, for example, that the target state is 
$$\bar{\rho}=\begin{bmatrix} 1 & 0 & 0 \\ 0 & 0 & 0 \\ 0 & 0 & 0 \end{bmatrix},$$
and that we dispose of two Lindblad generators, respectively associated to the following noise operators:
$$L_1=\begin{bmatrix} 0 & 1 & 0 \\ 0 & 0 & 0 \\ 0 & 0 & 0 \end{bmatrix} \quad L_2=\begin{bmatrix} 0 & 0 & 0 \\ 0 & 0 & 1 \\ 0 & 0 & 0 \end{bmatrix}.$$
It is easy to see that they satisfy Assunption 1: any non-extremal convex combination is in fact stabilizing -- this can be verified easily with the methods of \cite{ticozzi-QDS,ticozzi-NV}.
If the estimated initial state is 
$$\hat{\rho}_0=\begin{bmatrix} 0 & 0 & 0 \\ 0 & 1 & 0 \\ 0 & 0 & 0 \end{bmatrix},$$
then the state-based algorithms will try to use only the first generator to reach the target faster -- in fact the second generator would leave the estimated state invariant. But if the actual state is instead
$$\rho_0=\begin{bmatrix} 0 & 0 & 0 \\ 0 & 0 & 0 \\ 0 & 0 & 1 \end{bmatrix},$$
then the first generator alone does not stabilize the target, as $\rho_0$ is an invariant state for it. \qed

This example proves that if the switching rule is designed for the wrong initial state, then the algorithm may fail to provide a stabilizing law. 
However, we next present a result that allows us to design state-based strategies that are robust with respect to variations of the {\em actual} initial state, as long as the {\em estimated} one is a full rank state. 

\begin{prop}
If a family of TPCP maps $\{{\cal E}_t\}_{t=0}^{+\infty}$ is such that \beq\label{eq:limE}
\lim_{t\rightarrow+\infty} {\cal E}_t (\hat{\rho}_0) = \bar{\rho},
\eeq 
then for any $\rho_0$ such that $\supp(\rho_0)\subseteq\supp(\hat\rho_0)$ we have that
\beq
\lim_{t\rightarrow+\infty} \Tr(\bar\Pi{\cal E}_t ({\rho}_0)) = 1,
\eeq
where $\bar\Pi$ is the orthogonal projector on $\supp(\bar\rho).$
\end{prop}
\proof
As $\mathcal{E}_t$ is linear and the limit \eqref{eq:limE} exists, then for any scalar $\lambda$ we have 
\beq
\lim_{t\rightarrow +\infty} {\cal E}_t (\lambda \hat\rho_0) = \lambda \, \bar{\rho}.
\eeq

Since $\supp(\rho_0)\subseteq\supp(\hat\rho_0)$ by hypothesis, for any such density operator $\rho_0$, there exists a $\lambda>0$ such that,
\beq
\rho_0\leq \lambda \hat\rho_0. \eeq
In particular, one can always take $\lambda$ to be the inverse of the smallest eigenvalue of $\hat\rho_0$. Since the evolution is positivity preserving, it preserves matrix ordering and we can then conclude that
\beq\label{intres}
\lim_{t\rightarrow +\infty} {\cal E}_t (\rho_0) \leq  \lambda \bar{\rho}.
\eeq
Using \eqref{intres} and the fact that the evolution is trace-preserving we have:
\[
\lim_{t\rightarrow+\infty} \Tr(\bar\Pi{\cal E}_t ({\rho}_0)) =\lim_{t\rightarrow+\infty} \Tr({\cal E}_t ({\rho}_0))= 1.
\]
\qed

Assume now $\bar{\rho}$ to be a pure state. Then $\bar\Pi=\bar\rho$ is a one-dimensional projector, and the only density operator for which $\lambda\bar{\rho}$ is an upper bound is $\bar{\rho}$ itself. Furthermore, the condition $\supp(\rho_0)\subseteq\supp(\hat\rho_0)$ is satisfied if $\hat\rho_0$ is assumed to be full rank. We thus have the following:
\begin{cor}
If a family of TPCP maps $\{{\cal E}_t\}_{t=0}^{+\infty}$ is such that \beq
\lim_{t\rightarrow+\infty} {\cal E}_t (\hat{\rho}_0) = \bar{\rho},
\eeq 
for a full-rank initial state $\hat{\rho}_0$ and a pure target state $\bar{\rho}$, then $\bar{\rho}$ is GAS, i.e. it is asymptotically reached independently of the initial state $\rho_0$. 
\end{cor}

It is worth noting that the same results still holds if we relax CP to simple positivity, or if we consider discrete-time evolutions \cite{bolognani-arxiv}.

These stability results immediately imply that if a state-based switching law is designed for a full-rank initial state $\hat{\rho}_0$ in order to prepare a pure target state $\bar{\rho}$, then the corresponding evolution makes $\bar{\rho}$ GAS, i.e. it is asymptotically reached independently of the initial state $\rho_0$. Analogous reasoning can be obviously applied to subspace preparation.

In designing the control law we can always consider the completely mixed initial state, as one would do in the absence of initial information, or, if we have a good estimate $\hat{\rho}_0$ which is not full rank, we can consider a perturbed version $\tilde{\rho}_0=(1-\epsilon)\hat{\rho}_0+\epsilon I/d,$ with $\epsilon$ arbitrarily small, in order to ensure robustness. 

We could expect that choosing a state-based switching rule, without exploiting the information on the initial state, would lead to a stabilizing law characterized by a convergence rate comparable with a time-based switching rule, for which the initial state is a priori supposed to be unknown. Nevertheless, in the numerical examples that follow, we have that it still converges to the target with better performances than those obtained by implementing a time-based switching law.

\section{Examples} \label{sec:app}

In the following we compare the performances of different Markov evolutions that prepare an entangled pure state of interest, in terms of both a natural Lyapunov pseudo-distance and the Euclidean distance in coherence-vector representation (equivalent to the Hilbert-Schmidt norm). The evolutions we compare for both examples are four:
\begin{enumerate}
\item A time-independent semigroup generator, associated to the stabilizing convex combination of Assumption \ref{ass1}, for which the target $\rho$ is known to be the unique equilibrium (denoted in the figures as {\em no-switch});
\item A cyclic, time-based switched evolution, such that a convex combination of the alternating generators is equivalent to time-independent generator above (i.e. they satisfy Assumption 1). For both examples, a convex combination with uniform coefficients is sufficient, and used. A minimal switching time $\Delta T$ is chosen, so that the evolution is stabilizing. The Lyapunov function \eqref{defp} associated to the stabilizing convex combination will be called the {\em natural Lyapunov function};
\item A state-based switching evolution, where {the generator is chosen at $\Delta T$ intervals} in order to guarantee the fastest decrease of the natural Lyapunov function, as described in Section \ref{sec:steepestdesc},  depending on a full-rank estimate of the initial state. The same evolution is also applied to a different initial state, in order to verify the robustness of the approach, and the potential deterioration in performances;
\item A state-based switching evolution, where the generator is chosen as described in Section \ref{sec:suboptimal} depending on a full-rank estimate of the initial state. The same evolution is also applied to a different initial state, in order to illustrate the robustness of the approach, and the potential deterioration in performances;
\end{enumerate}
Choosing a pure target allows us to compare the performance of state-based switching strategies with correct initializations and faulty ones, since the results of Section \ref{sec:robustness} guarantee convergence. Using a distance that is not tailored to the problem (the Euclidean one) help us to illustrate how the convergence in Lyapunov distance does not hides undesired behaviors. For all simulations, we use an integration step (the {time unit} for the X axis of the plots) of $t=0.02,$ and a minimal switching interval $\Delta T=3\,t.$

\subsection{Bell states}

We first consider a two-qubit system defined on ${\cal H}_{AB}={\cal H}_A\otimes{\cal H}_B\sim \textrm{span}\{\ket{0},\ket{1}\}^{\otimes 2}$. Our aim is to prepare the maximally entangled Bell state,
\beq
\bar{\rho}_{AB}=\frac{1}{2}(\ket{00}+\ket{11})(\bra{00}+\bra{11}),
\eeq
by switching between Lindblad dynamics and to compare time-based and state-based switching rules convergence rates. Denote as usual the Pauli matrices as:
\beq
\sigma_x=\begin{bmatrix} 0 & 1 \\ 1 & 0 \end{bmatrix},\quad\sigma_y=\begin{bmatrix} 0 & i \\ -i & 0 \end{bmatrix},\quad
\sigma_z=\begin{bmatrix} 1 & 0 \\ 0 & -1 \end{bmatrix},
\eeq
with the scaled identity $\sigma_0=\frac{1}{\sqrt{2}}I$.

The first generator we consider is associated to the Hamiltonian $H=\sigma_y\otimes\sigma_0+\sigma_0\otimes\sigma_y$. The second generator is determined by the Lindblad operator $L=\sigma_z\otimes\sigma_0-i(\sigma_y\otimes\sigma_x)$. The target state $\bar{\rho}_{AB}$ is a fixed state for both the generators and it can be proved  that applying them simultaneously, with any nonzero weight, makes $\bar{\rho}_{AB}$ asymptotically stable \cite{ticozzi-markovian}. 

Let us denote $\hat{L}_1$ and $\hat{L}_2$ the superoperator form of the Lindblad generators derived by $H$ and $L$ respectively, according to the coherence-vector representation, as in \ref{sdst}. 
After applying the change of basis that makes the generators of the reduced dynamics linear, it is easy to numerically check that $A_c=\frac{1}{2}A_1+\frac{1}{2}A_2$ is actually a Hurwitz convex combination. This combination satisfies Assumption 1 and allows us to derive time-based and state-based switching rules that prepare the target state as described in the introduction to this section. 

\begin{figure}[!h]
  \centering
  \includegraphics[width=9cm]{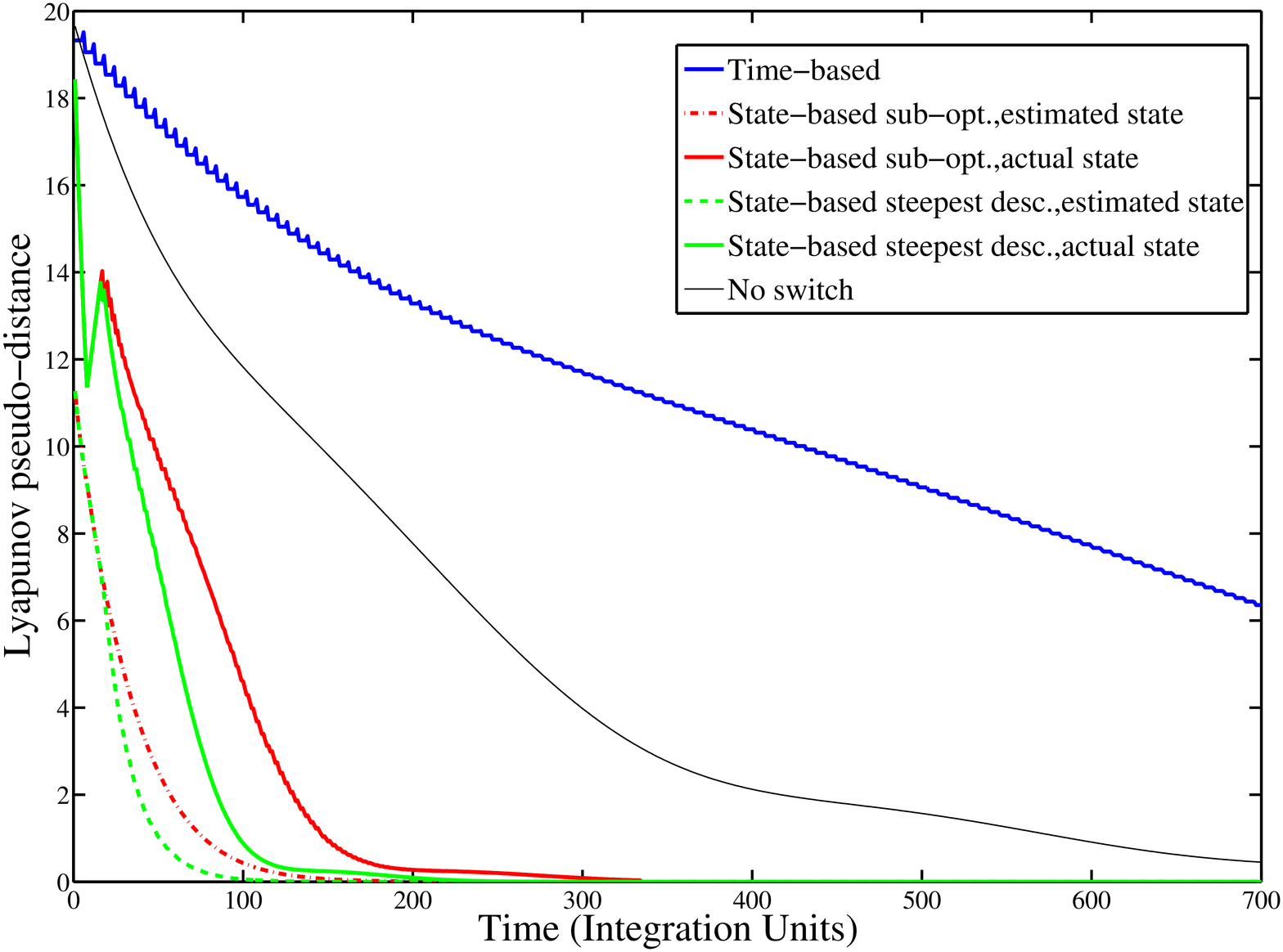}
  \caption{(Color online) {Evolution the Lyapunov function with different switching rules in the two-qubit case.}}
  \label{fig:bell1}
\end{figure}

\begin{figure}[!h]
  \includegraphics[width=9cm]{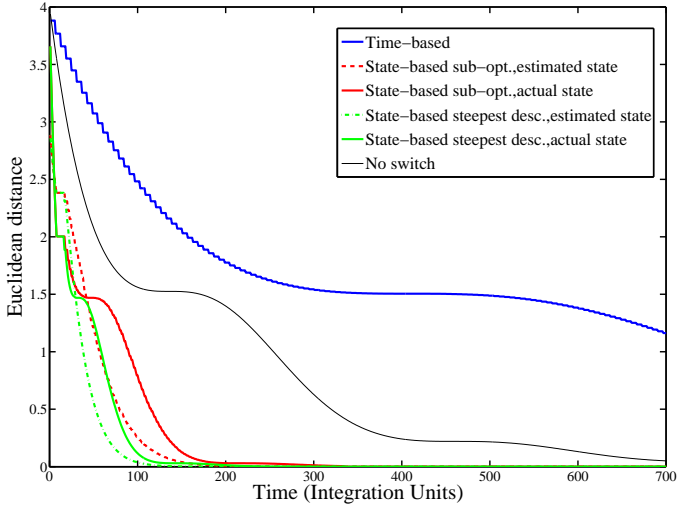}
  \caption{(Color online) {Evolution the Euclidean distance with different switching rules in the two-qubit case.}}
  \label{fig:bell2}
\end{figure}

In Figure \ref{fig:bell1} and \ref{fig:bell2} we compare the convergence features in terms of the natural Lyapunov function, and in terms of the coherence-vector euclidean distance to the target, respectively. The actual initial state is $\rho=\ket{0}\bra{0}\otimes\ket{0}\bra{0}$, while the estimated initial state is chosen to be the completely mixed one. The estimated state is used to compute the state-based switching strategies, whose estimated behavior is depicted with dashed lines. All the solid lines represent the evolution of the distances calculated for the actual states. We set the parameters for the suboptimal state-based switching to be $r_1=r_2=1$.

In Fig.\ref{fig:bell1} we find, as expected, that the time-based switching is the slowest to converge, and at close inspection it even exhibits local increase in the Lyapunov function. In fact, time-based switching is an approximation of the no-switching strategy, and during some switching steps the state may also tend to get farther from the target -- it is only after a full cycle that we are guaranteed, for sufficiently fast switching, that we the distance to the target has decreased. In our simulations, the bad performance is caused by the fairly large cycle time, making the cyclic switching a rough approximation of the reference convex combination. In this particular case, the effect is likely strengthened by the presence of a purely Hamiltonian generator among the switching ones, which does not induce dissipative, and hence contractive, effects. 

The convex-combination, Hurwitz generator displays a smooth converging behavior, with the distance from the target state with the state-based rule is monotonically decreasing. The state-based strategies show, for both the estimated and the actual state behavior, faster convergence. The evolution of the actual state converges to the target state but it is not monotone, as it is clear by looking at the peaks around 50 time units. Nonetheless, it remains much more effective than the evolution associated to convex combination of Assumption \ref{ass1}, where no switching is performed and all generators are activated at each time scaled by the corresponding weights, for both state-based strategies. The ``steepest descent'' approach is, as expected, overall slightly faster than the suboptimal one with guaranteed minimal descent. 

The qualitative evaluation of the plots of Fig.\ref{fig:bell2} leads to the same conclusions, confirming the advantage of the state-based strategies, even in presence of a wrong initialization of the algorithm. In fact, the initial convergence appears to be faster for the actual state. 

\subsection{GHZ states}

We here present an example of preparation of a three-qubit GHZ state, 
\beq
\bar{\rho}_{GHZ}=\frac{1}{2}(\ket{000}+\ket{111})(\bra{000}+\bra{111}),
\eeq
with the switching techniques we introduced above. The stabilizing, non switching generator we consider is one that has been proposed in \cite{ticozzi-steady}, and can be implemented via  an Hamiltonian \[H=\sigma_x^{(1)}-\sigma_x^{(2)}\otimes\sigma_x^{(3)},\] 
where $\sigma_x^{(i)}=I\otimes\dots\otimes\sigma_x\otimes\dots\otimes I$, and $\sigma_x$ acts on the $i$-th qubit, along with two noise operators, \[L_1=(\ket{00}\bra{01}+\ket{11}\bra{10})\otimes I,\]
\[L_2=I\otimes(\ket{00}\bra{01}+i\ket{11}\bra{10}).\] 
The simultaneous action of the three components, with arbitrary positive weights, leads to the asymptotic stabilization of $\bar{\rho}_{GHZ}$. A stabilizing time-based switching rule can be therefore easily obtained with a fast periodical switching between them. We then consider switching between ${\cal L}_{0,1,2},$ corresponding to the superoperator action of $H, L_1, L_2,$ respectively. 

The actual initial state is $\rho=\ket{0}\bra{0}\otimes\ket{0}\bra{0}\otimes\ket{0}\bra{0}$, while the estimated initial state is chosen to be the completely mixed one. The estimated state is used to compute the state-based switching strategies, whose estimated behavior is depicted with dashed lines. All the solid lines represent the evolution of the distances calculated for the actual states. We set the parameters for the suboptimal state-based switching to be $r_1=r_2=r_3=1$.

\begin{figure}[!h]
  \centering
  \includegraphics[width=9cm]{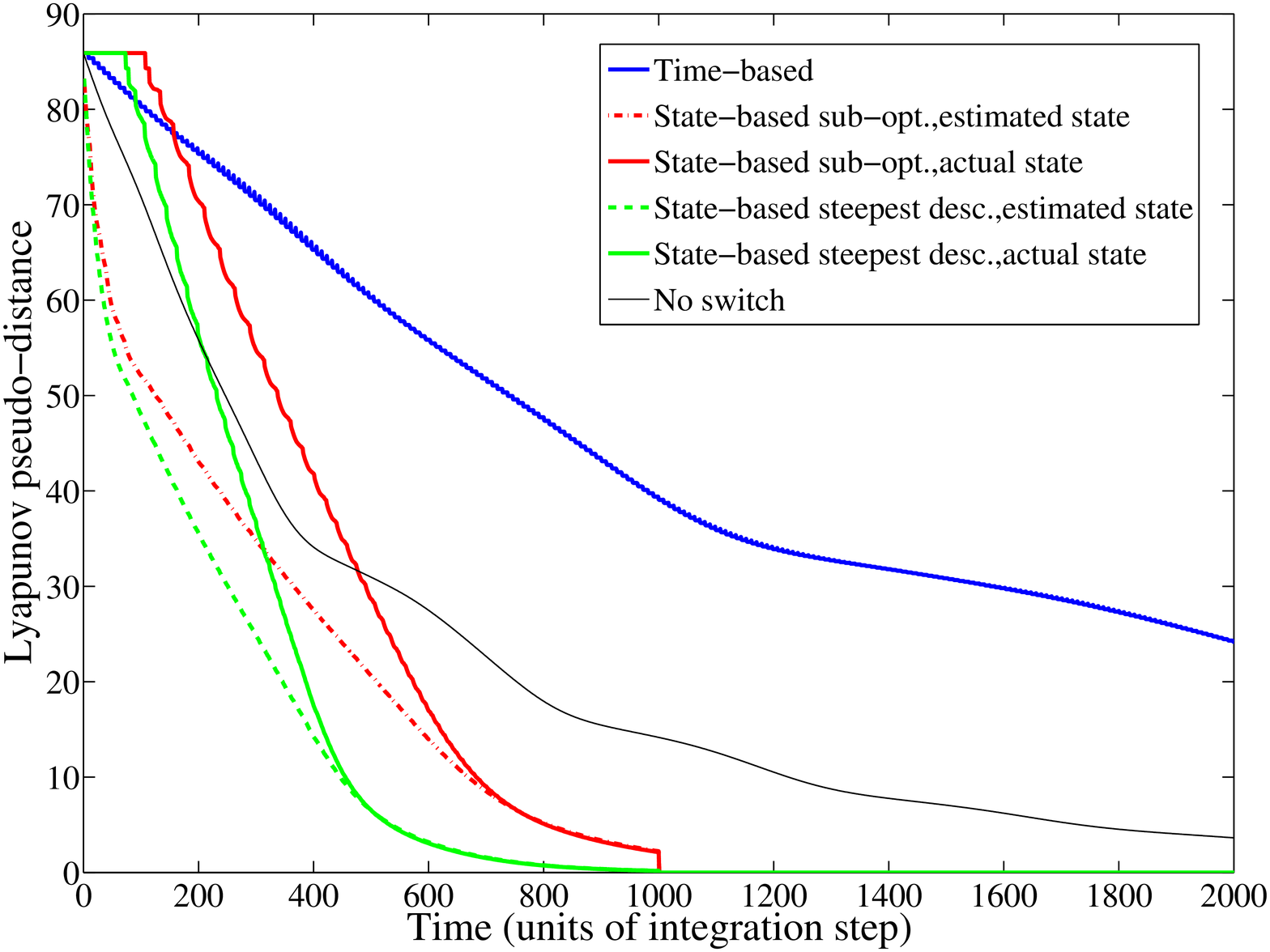}
  \caption{(Color online) {Evolution the Lyapunov function with different switching rules for the three-qubit case.} }
  \label{fig:ghz1}
\end{figure} 

\begin{figure}[!h]
  \centering
  \includegraphics[width=9cm]{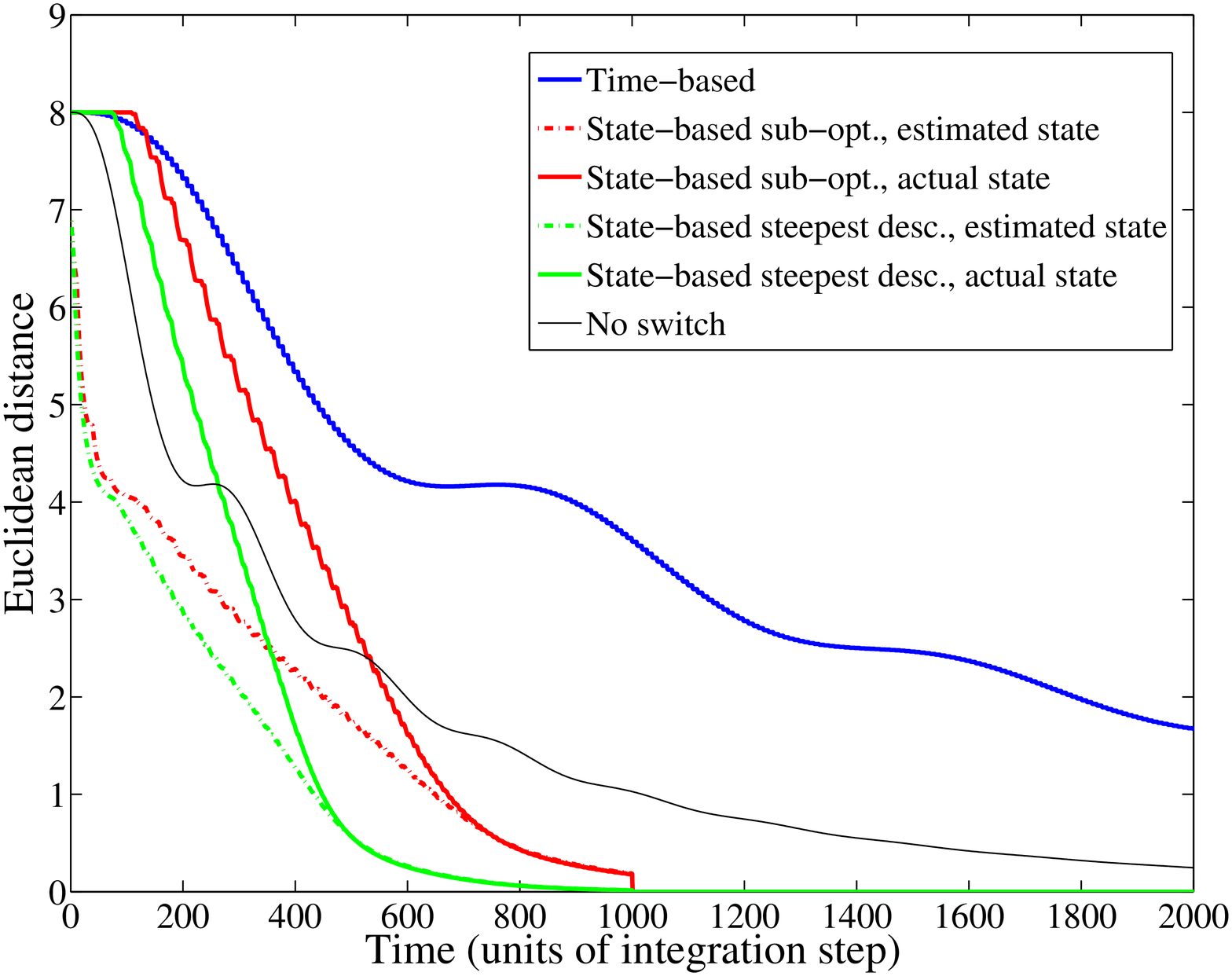}
  \caption{(Color online) {Evolution the Euclidean distance with different switching rules for the three-qubit case.}}
  \label{fig:ghz2}
\end{figure} 

In Fig.\ref{fig:ghz1} we find, as expected, that the time-based switching is the slowest to converge in terms of the Lyapunov function, and close inspection reveals that the cycling indices some local fluctuations. This is because of the fairly large cycle time, making the time-based switching a rough approximation of the reference convex combination. The effect may also be partially due to the presence of a purely Hamiltonian generator that does not induce dissipative contraction effects. The non-switching generator associated to the stabilizing convex combination displays a smooth, monotonically decreasing behavior, as expected. The state-based strategies show, for both the estimated and the actual state behavior, the fastest overall convergence. The evolution of the actual state converges to the target state but the Lyapunov pseudo-distance does not start decreasing immediately, a clear sign that the optimal generator for the completely mixed state is not effective for the actual initial state. Nonetheless, after another generator is selected, state-based strategies become more effective than the evolution associated to the convex combination of Assumption \ref{ass1}, without switching. The ``steepest descent'' approach is overall slightly faster than the suboptimal one with guaranteed minimal descent. 

A careful scrutiny of the plots of Fig.\ref{fig:ghz2}, for the undistorted euclidean distance in coherence vector space, essentially leads to the same conclusions, confirming the advantage of the state-based strategies, even in presence of an incorrect initialization of the algorithm. 

\section{Discussion and Perspective}
In this paper we have presented two approaches for designing switching quantum Markov evolutions that make a desired state GAS. For algorithms obtained by using either time-based or state-based designs, the underlying assumption is the existence of a convex combination of the switchable generators that would make the target GAS. The first approach we propose constructs time-based, cyclic switching laws: its advantages are the simplicity, and the intrinsic robustness with respect to the initial state. No initial state estimate, optimization or on-line computation is needed. However, being essentially a way to approximate the stabilizing convex combination, the method is quite sensitive to the minimum allowed switching time. Numerical simulations, as exemplified by those presented in this paper, show that this strategy is in general the worst performer of the group.

State-based strategies, while they entail some computational overhead, guarantee to outperform the non-switching evolution when the initial state is accurately known, and numerical simulations show that they retain this advantage even with a generic initialization (e.g. the completely mixed state we used). Robustness with respect to the initial state can be guaranteed if the target is a pure state (or a subspace) by arbitrarily small perturbation of the initial estimate. 
The state-based strategies thus offer a valid alternative to time-invariant methods for dissipative preparation of states, especially when implementation of complex dynamics in an experimental setting is challenging, and further motivates their study. 

The structure of the algorithm suggests that better results, in terms of robustness, may be attained by combining optimal state-based switching with state-estimation methods. The latter could in principle be implemented via  continuous measurements and filtering equations, when these guarantee asymptotic convergence of the estimated state to the actual one \cite{vanhandel-phd,rouchon-filtering,ticozzi-stochastic}. It is known that, for pure states and subspaces with time-independent stochastic dynamics, convergence in expectation implies convergence in probability \cite{ticozzi-stochastic}. An extension of these results would help proving the effectiveness of the filtering-based switching. This combination would lead to a control strategy that closely resembles a classical one, where state-based switching is typically decided via evaluation of the current state through a feedback loop. 

Other potential developments of these results may include a comparison with the performance of optimal control methods for state preparation and integration with scalable strategies for the dissipative preparation of entangled states on large networks \cite{ticozzi-ql,ticozzi-steady}.
 
\acknowledgments
F.T. thanks Lorenza Viola  for fruitful conversations on the topics of this work and Maria Elena Valcher for suggesting some relevant references. F. T. acknowledges hospitality from the Physics and Astronomy Department at Dartmouth College -- where part of this work was performed, and support by the QUINTET and QFuture projects of the University of Padova. 

\appendix
\section{Proof of Proposition \ref{prop:linswitch}}\label{app:proof1}
\proof
{\ft Consider a periodic switching with period $\varepsilon$. During each period, each system $A_j$ acts for a time interval proportional to its coefficient $\alpha_j$ (the order is not important, as we shall see). The transition matrix over a period can be expressed as the exponential of an {\em effective} generator $\bar A$: 
\begin{equation*}
\exp(\bar{A}\varepsilon)=\exp(\alpha_mA_m\varepsilon)\hdots \exp(\alpha_1A_1\varepsilon).
\end{equation*}
We can then separate the contribution of the {\em average} generator $A_c$ with respect to the part depending on $\varepsilon$ as $\bar{A}:=A_c+\Upsilon_c\varepsilon$ and $\Upsilon_c$ is bounded and contains all the higher-order terms of a Taylor expansion of $\bar A$ around $\varepsilon=0$.} The eigenvalues of a matrix continuously depend on its entries. As $\Upsilon_c$ is bounded, if $\varepsilon\rightarrow 0$, then the eigenvalues of $\tilde{A}$ approach those of $A_c$ and assuming that $A_c$ is Hurwitz, there exists a $\varepsilon>0$ for which $\tilde{A}$ is Hurwitz too.  Fixed such an $\varepsilon$, a periodic switching path can be defined e.g. as follows:
\begin{equation}
\label{stabilswitch}
\sigma(t)=\begin{cases} 1 & \textrm{if} \mod(t,\varepsilon)\in[0,\alpha_1\varepsilon) \\
2 & \textrm{if} \mod(t,\varepsilon)\in[\alpha_1\varepsilon,\alpha_1\varepsilon+\alpha_2\varepsilon) \\ \vdots & \\ m &  \textrm{if} \mod(t,\varepsilon)\in[(\sum_{j=1}^{m-1}\alpha_j)\varepsilon,\varepsilon)\end{cases}
\end{equation}
Setting $s_1\le s_2,$ let us define
\begin{equation*}
\phi(s_2,s_1):=e^{A_p(s_2-t_p)}e^{\alpha_{p-1}A_{p-1}\varepsilon}\hdots e^{A_{k-1}(t_k-s_1)},
\end{equation*}
the transition matrix from the state at instant $s_1\in(t_{k-1},t_k)$ to that at instant $s_2\in(t_p,t_{p+1})$ according to a given switching law. For any non-negative integers $l_1\le l_2$ the evoution covers a finite number of cycles, so that
\begin{equation*}
\phi(l_2\varepsilon,l_1\varepsilon)=e^{\bar{A}(l_2-l_1)\varepsilon}.
\end{equation*}
As $\bar{A}$ is Hurwitz, there exist positive numbers $\kappa$ and $\lambda$ such that
\begin{equation*}
||\phi(l_2\varepsilon,l_1\varepsilon)||\le\kappa e^{-\lambda(l_2-l_1)\varepsilon}.
\end{equation*}
For any $s_1\le s_2$, let $l_1$ and $l_2$ satisfy
$$l_1\varepsilon\le s_1<(l_1+1)\varepsilon, \quad (l_2-1)\varepsilon<s_2\le l_2\varepsilon.$$
Then we have:
\begin{eqnarray*}
||\phi(s_2,&&\hspace{-3.5mm}s_1)||\le||\phi(l_1\varepsilon,s_1)||\cdot||\phi(l_2\varepsilon,l_1\varepsilon)||\cdot||\phi(s_2,l_2\varepsilon)||\\
&&\hspace{-5mm}\le\kappa e^{-\lambda(l_2\varepsilon-l_1\varepsilon)}||\phi(0,s_1-l_1\varepsilon)||\cdot||\phi(0,l_2\varepsilon-s_2)||.
\end{eqnarray*}
Finally, by denoting:
$$\kappa_1=\max_{0\le t\le\varepsilon}{||\phi(0,t)||},$$
which is always attainable because $\phi(0,t)$ is continuous in $t$, we get 
$$||\phi(s_2,s_1)||\le\kappa_1^2\kappa e^{-\lambda(l_2-l_1)\varepsilon} \le\kappa_1^2\kappa e^{-\lambda(s_2-s_1)}.$$
The transition matrix is exponentially convergent and the switching system is then stabilized.\qed


\section{Proof of Theorem \ref{thm:suboptimal}}\label{app:proof2}
\proof
{\ft We want to show now that such a switching law is acceptable, that is actually $t_{k+1}>t_k$. Let $\theta$ be any real number greater than 1. 
We first consider the case:
\begin{equation}
\label{case1}
||x(t)||\le\theta||x_{k+1}|| \quad\forall t\in[t_k,t_{k+1}].
\end{equation}
and define 
\begin{equation*}
g(t)=x(t)^T(Q_j+I)x(t) \quad t\in[t_k,t_{k+1}],
\end{equation*}
where $j=\sigma(t_{k^+})$. On the one hand, thanks to Lemma \ref{lemmaprec} we know that
\begin{equation*}
x(t_k)^TQ_jx(t_k)\le -x(t_k)^Tx(t_k),
\end{equation*}
and hence we are guaranteed that
\begin{equation}
\label{aqwe}
g(t_k)\le0.
\end{equation}
On the other hand, from \eqref{item1},we have
\begin{equation}
\label{awer}
g(t_{k+1})\ge(1-r_j)x_{k+1}^Tx_{k+1}.
\end{equation}
Deriving the latter we get
\begin{equation*}
\frac{dg}{dt}=x(t)^T(A_j^T(Q_j+I)+(Q_j+I)A_j)x(t).
\end{equation*}
By denoting
\begin{equation*}
\eta_j:=||A_j^T(Q_j+I)+(Q_j+I)A_j||,
\end{equation*}
and using \eqref{case1}, we have
\begin{equation}
\label{polk}
|\frac{dg}{dt}|\le\theta^2\eta_jx_{k+1}^Tx_{k+1} \quad \forall t\in[t_k,t_k+1].
\end{equation}
According to \eqref{aqwe} and \eqref{awer}
\begin{equation*}
\frac{g(t_{k+1})-g(t_k)}{t_{k+1}-t_k}\ge\frac{(1-r_j)x_{k+1}^Tx_{k+1}}{t_{k+1}-t_k}.
\end{equation*}
Remembering then \eqref{polk},
\begin{equation*}
\frac{(1-r_j)x_{k+1}^Tx_{k+1}}{t_{k+1}-t_k}\le\theta^2\eta_jx_{k+1}^Tx_{k+1},
\end{equation*}
and, consequently,
\begin{equation*}
t_{k+1}-t_k\ge\frac{(1-r_j)}{\theta^2\eta_j}.
\end{equation*}
This shows that the switching instants are have minimal distance between them, and there cannot be infinite switching times in any finite time interval.

We are then left with the other case, when \eqref{case1} does not hold, and thus:
\begin{equation*}
\exists t^\ast\in[t_k,t_{k+1}):  ||x(t^\ast)||>\theta||x_{k+1}||.
\end{equation*}
As the system dynamics in this time interval is described by $A_j$, then
\begin{equation*}
x(t^\ast)=\exp(A_p(t^\ast-t_{k+1}))x_{k+1}.
\end{equation*}
From the latter, and remembering that 
\begin{equation*}
||\exp(A_j(t^\ast-t_{k+1}))||\le \exp(||A_j||(t_{k+1}-t^\ast)),
\end{equation*}
it follows that a minimal spacing is also guaranteed:
\begin{equation*}
t_{k+1}-t_k\ge t_{k+1}-t^\ast>\frac{\ln(\theta)}{||A_j||}.
\end{equation*}
Finally, gathering both the cases, we can say that
\begin{equation*}
t_{k+1}-t_k\ge \sup_{\theta>1}\min_{j\in\mathcal{P}}\biggl(\frac{1-r_j}{\theta^2\eta_j},\frac{\ln(\theta)}{||A_j||}\biggr),
\end{equation*}
and that the switching signal is valid as the difference is always positive. 

Choosing then $V(x)=x^TPx$ as Lyapunov function we get
\begin{equation*}
\frac{dV}{dt}=x^T(t)Q_{\sigma(t)}x(t)\le r_{\sigma(t)}x^T(t)x(t)\le-rx^T(t)x(t),
\end{equation*}
where 
\begin{equation*}
r:=\min\{r_1,\hdots,r_m\},
\end{equation*}
and the theorem is proved.
\qed }

\bibliographystyle{apsrev}
\bibliographystyle{plain}
\bibliography{bib-purification-2}

\end{document}